\documentclass[aps,prb,showpacs,twocolumn,superscriptaddress]{revtex4}
\usepackage{graphicx}
\usepackage{bm}
\usepackage{amssymb}
\usepackage[usenames]{color}

\def\rar{\rightarrow}
\def\oH{\hat H}
\def\oc{\hat c^{\phantom\dagger}}
\def\ocd{\hat c^\dagger}
\def\on{\hat n}
\def\ua{\uparrow}
\def\da{\downarrow}
\def\bA{{\bf A}}
\def\bE{{\bf E}}
\def\bR{{\bf R}}
\def\bk{{\bf k}}
\def\e{\varepsilon}
\def\m[#1]{\phantom{.}\underline{\underline{#1}}}

\def\p{\partial}
\def\rp{\right)}
\def\lp{\left(}
\def\ra{\rangle}
\def\la{\langle}

\def\bare{{\bar\varepsilon}}

\def\bars{{\bar\sigma}}

\def\fff#1/#2{\leavevmode\kern.1em\raise.5ex\hbox{\the\scriptfont0 #1}\kern-.1em/\kern-.15em\lower.25ex\hbox{\the\scriptfont0 #2}}

\begin{document}

\title{Long-lived nonequilibrium states in the Hubbard model with an electric field}

\author{Alexander V. Joura}
\email{Alexander.Joura@physnet.uni-hamburg.de}
\affiliation{Institute of Theoretical Physics, University of Hamburg, Jungiusstrasse 9, D-20355 Hamburg, Germany}

\author{J.~K.~Freericks}
\affiliation{Department of Physics, Georgetown University, Washington, DC 20057-0995, USA}

\author{Alexander I. Lichtenstein}
\affiliation{Institute of Theoretical Physics, University of Hamburg, Jungiusstrasse 9, D-20355 Hamburg, Germany}


\date{\today}

\begin{abstract}
We study the single-band Hubbard model in the presence of a large spatially uniform electric field out
of equilibrium. Using the Keldysh nonequilibrium formalism, we solve the problem using perturbation
theory in the Coulomb interaction $U$. We present numerical results for the charge current,
the total energy of the system and the double occupancy on an infinite-dimensional hypercubic lattice with nearest-neighbor hopping. The system is isolated from an external bath and is in the paramagnetic state. We show that an electric
field pulse can drive the system to a steady nonequilibrium
state, which does not evolve into a thermal state. We compare results obtained within second-order perturbation theory (SOPT),
self-consistent second-order perturbation theory (SCSOPT) and iterated perturbation theory (IPT). We also discuss the
importance of initial conditions for a system which is not coupled to an external bath.
\end{abstract}

\maketitle

\section{Introduction} 
The recent growth of experimental interest in systems driven out of equilibrium~\cite{pprobe} has stimulated a lot of theoretical activity to explain these experiments.
The observation of Bloch oscillations in ultracold atoms~\cite{cold} and the corresponding theoretical treatment
of the fermionic Hubbard model in an additional linear potential~\cite{rosch} give a new interesting twist to
the classical problem of the electric current on a lattice and require nonequilibrium many-body methods,
developed by Kadanoff and Baym~\cite{kadbaym}, and Keldysh~\cite{kelda}.

The nonequilibrium dynamical mean-field theory (NE-DMFT), introduced by Schmidt
and Monien~\cite{schmidt.monien} and Freericks, {\it et al.}~\cite{ftz}, is a combination of the
equilibrium~\cite{dmft} and nonequilibrium DMFT formalisms. It has made
feasible nonperturbative calculations of many-body models driven out of equilibrium by external
fields. While for the Falicov-Kimball (FK) model it was possible to formulate a definitive
numerically exact DMFT algorithm~\cite{ftz,jkf.trans}, for the Hubbard model, one currently has
to choose between quantum Monte Carlo methods and perturbative calculations.
The nonequilibrium continuous time quantum Monte Carlo (NE CT-QMC) technique~\cite{noneq-ctqmc,tsuji.ctqmc} is a
powerful calculational tool, but it suffers from the so-called phase problem, which renders its usage to relatively short real times. Hence, most Hubbard model results for NE-DMFT are perturbative.

Recently, there has been a significant interest in the many-body thermalization of isolated systems~\cite{rigol}. Cramer, {\it et. al.}~\cite{cramer} studied the Bose-Hubbard model and provided general arguments
that after the interaction quench to a noninteracting state the system will relax to a nonthermal state.
The opposite case of an interaction quench to a nonzero interaction, was studied by Eckstein and Kollar~\cite{eck-quench2-prl}
for the Falicov-Kimball model on the Bethe lattice using a numerically exact NE-DMFT approach. It was found that the system never
thermalizes. The inability of the system to thermalize was attributed to the presence of an infinite number of conserved quantities for the Falicov-Kimball model,
due to the presence of immobile $f$-particles. Later Eckstein, {\it  et al.}~\cite{eck-quench-prl}
found that the Hubbard model directly thermalizes only for interactions close to $U_{c}^{dyn}$. For quenches to
interactions both smaller and larger, the system is initially trapped in a quasistationary nonthermal state,
called a prethermalized state. Using the quantum Bolzmann equation, Moeckel and Kehrein~\cite{mokel} argued
that for small values of the Coulomb interaction these prethermalized states eventually evolve into thermal states on timescales
of the order of $\tau_{therm}\sim \rho_0^{-3}(0)U^{-4}$.

Despite the fact that the prethermalized states cannot be described by a simple Fermi
distribution, interaction quenches in the Hubbard model usually do lead to distributions qualitatively similar to Fermi
ones. In particular the double occupancy in prethermalized states decreases as the repulsive interaction increases.
An interesting change in behavior happens when, instead of quenching the interaction, one quenches an external
electric field. When a DC field is applied to a system, it is predicted to heat up~\cite{prelov} to $T=\infty$ as the current
goes to zero, thus providing an example of thermalization. Fotso, {\it et al.}~\cite{fotso}, predict that this is just one of five
different scenarios that can occur for such a field quench.
Tsuji, {\it et. al.}~\cite{tsu-pulse} studied the Hubbard model under the application of an electric pulse
to the system, which is initially prepared in an interacting thermal state (no $U$-quench).  They found that by tuning
the pulse parameters it is possible to achieve a long-lived state, corresponding to a
thermal state with a negative temperature, where electrons behave as though they attract to each other.

In this article, we study the behaviour of the Hubbard model in the presence of a uniform
electric field. In particular, we show how a combination of the interaction quench and an electric field pulse
can drive the system to a long-lived nonequilibrium state, where the particle distribution
does not resemble a Fermi distribution. Our approximate calculations show that the system conserves
its exotic properties even for times comparable to theoretical estimates of the lifetime of prethermalized
states. In Sec.~II, we develop the formalism, in Sec. III, we show the numerical results, and
in Sec.~IV, we present our conclusions.

\section{Formalism} 
We consider the single-band Hubbard model on a $d$-dimensional Bravais lattice with
the Hamiltonian
\begin{equation}\label{ham}
\oH(t) = \sum_{ij,\sigma} T_{ij}(t) \ocd_{i\sigma} \oc_{j\sigma} -\mu_0\sum_{i\sigma} \ocd_{i\sigma} \oc_{i\sigma}
+ U(t)\sum_i \on_{i\ua} \on_{i\da}
\end{equation}
where $T_{ij}(t)$ are hopping coefficients with the time-dependence described below, $\mu_0$ is the noninteracting chemical potential and $U(t)$ is
the on-site Coulomb interaction (which can also be
time-dependent). We describe the external spatially uniform electric field via the vector potential $\bA(t)$
\begin{equation}
\bE(t) = -\p_t \bA(t) \ .
\end{equation}
The Peierls substitution~\cite{peisub} is used to account for the electric field in the Hamiltonian, so the hopping
matrix elements satisfy
\begin{equation}\label{psub}
T_{ij}(t) = T_{ij} \exp [ -i \bA(t)\cdot (\bR_j-\bR_i) ] \ .
\end{equation}
In $\bk$-space, the noninteracting part of the Hamiltonian in Eqs.~(\ref{ham}) and (\ref{psub}) becomes diagonal
\begin{equation}\label{hok}
\oH_0(t) = \sum_\bk \xi(\bk -\bA(t)) \ocd_{\bk\sigma} \oc_{\bk\sigma}
\end{equation}
where $\xi(\bk) =\e(\bk)-\mu_0$ and $\e(\bk)$ is the dispersion law
$\e(\bk) = \sum_{i} T_{0i}e^{i\bk \cdot \bR_i}$.

In order to investigate the Hamiltonian in Eq.~(\ref{ham}), we use the Keldysh nonequilibrium
Green's function formalism. For the details of the formalism, we refer the reader to
the original article~\cite{kelda} and the review by Rammer
and Smith~\cite{rams}. At time $t_0=0$, the system is prepared in thermal equilibrium at
temperature $T_0$ with $E(t_0) = U(t_0)=0$.
Then one can study various profiles of turning on $U(t)$ and $E(t)$.

The matrix Green's function is expressed using creation/annihilation operators in the Heisenberg representation as
\begin{equation}\label{magr}
\m[G]_{\bk\sigma}(t,t') =G_{\bk\sigma}^{\alpha\beta}(t,t') = -i\la T_C \ \!\oc_{\bk\sigma}(t_\alpha) \ocd_{\bk\sigma}
(t'_\beta) \ra_0
\end{equation}
where $T_C$ is the time-ordering operator along the Keldysh contour; indices $\alpha,\beta=\pm$, determine
whether the corresponding time lies on the forward or return branch of the Keldysh contour; and $\la ... \ra_0$
denotes the thermal average with respect to the initial noninteracting thermal density matrix
\begin{equation}
\la... \ra_0 = \frac{{\rm Tr}(\hat\rho_0 ...)}{{\rm Tr}(\hat\rho_0)} \ , \qquad \hat\rho_0 = e^{-\hat H(t_0)/T_0}
\end{equation}
with $T_0$ being the temperature at $t_0$ (and the vector potential vanishes at $t_0$). Analytic formulas for the noninteracting Green's functions
$\m[G]_{\bk\sigma}^{0}(t,t')$ have been derived in Refs.~\onlinecite{jauhowilk} and \onlinecite{jkf0}.
We would like to emphasize that those are exact solutions for the Hamiltonian in Eq.~(\ref{hok}), which means that
the electric field is taken into account {\it non-perturbatively}. So in the following by the term
``noninteracting'' we mean functions where the electric field is included, but the interaction between
electrons is not.

The matrix Green's function in Eq.~(\ref{magr}) obeys the Dyson equation
\begin{equation}
\m[G]_{\bk\sigma} = \m[G]_{0,\bk\sigma} + \m[G]_{0,\bk\sigma} \otimes \m[\Sigma]_{\bk\sigma} \otimes \m[G]_{\bk\sigma}
\label{dmat}
\end{equation}
where the symbol $\otimes$ denotes the time convolution and matrix multiplication
\begin{equation}\label{convol}
(\m[A] \otimes \m[B])^{\alpha\beta}(t,t') = \int\limits_{t_0}^{+\infty} \sum_\gamma A^{\alpha\gamma}(t,t_1) B^{\gamma\beta}(t_1,t') dt_1.
\end{equation}

For the Hubbard Hamiltonian in Eq.~(\ref{hok}), the charge current density operator is
\begin{equation}\label{jop}
\hat j_{\alpha\sigma}(t) = \frac{e}{\hbar} \sum_\bk \p_{k_\alpha} \xi [ \bk-\bA(t) ] \ocd_{\bk\sigma} \oc_{\bk\sigma}
\end{equation}
and the current density can be found using the equal-time lesser Green's function~\cite{jkf0} $G^<=G^{+-}$
\begin{equation}\label{jcur}
j_{\alpha\sigma}(t) = -i \sum_\bk G^{<}_{\bk\sigma}(t,t) \frac{\p}{\p k_\alpha} \e [\bk-\bA(t)].
\end{equation}
The lesser Green's function $G^{<}_{\bk\sigma}(t,t')$ also allows one to calculate the thermal
average of the double occupancy operator $\hat D = \sum_i \on_{i\sigma}\on_{i\bars}$
\begin{equation}\label{double}
D(t) = \frac{i}{U(t)}\sum_\bk \lim\limits_{t'\rar t}\big\{-i\p_t-\mu_0+\e\big[\bk-\bA(t)\big]\big\}
G^{<}_{\bk\sigma}(t,t'),
\end{equation}
and the total energy of the system satisfies
\begin{equation}\label{etot}
E_{tot}(t) = i\sum_\bk \lim\limits_{t'\rar t}\big\{-i\p_t-\mu_0-\e\big[\bk-\bA(t)\big]\big\}
G^{<}_{\bk\sigma}(t,t').
\end{equation}

\begin{figure}[!htb]
\begin{center}
\includegraphics[width=.3\linewidth]{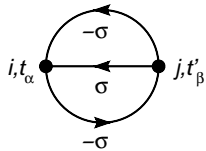}
\end{center}
\caption{Second-order contribution to the self-energy. The solid dot vertices correspond to the interaction $U$ and time
changes along the horizontal direction.  The solid lines represent either the Hartree-Fock (SOPT),
or the dressed Green's function (SCSOPT), or the effective medium for the equivalent impurity problem (IPT).}
\label{secd}
\end{figure}

When the shift $U(t) n_{\bars}$ of the chemical potential is incorporated into the noninteracting Green's
functions ({\it i.~e.}~perturbation theory is implemented in terms of the Hartree-Fock Green's functions), we have
only a single diagram for the second-order contribution to the self-energy shown in
Fig.~\ref{secd}.

When the lines correspond to the Hartree-Fock Green's functions, we obtain the second-order perturbation
theory (SOPT). If we use the dressed Green's functions for the lines, we obtain the self-consistent second-order
perturbation theory (SCSOPT). Thus for the SCSOPT case the corresponding formula is
\begin{equation}\label{sela2}
\Sigma_{ij,\sigma}^{\alpha\beta (2)}(t,t') = U(t) G_{ij,\sigma}^{\alpha\beta}(t,t')
G_{ji,\bars}^{\beta\alpha}(t',t) G_{ij,\bars}^{\alpha\beta}(t,t') U(t').
\end{equation}

We use the limit of infinite dimensions on the hypercubic lattice with nearest neighbor ($nn$) hopping,
introduced in Ref.~\onlinecite{mvprl}, which simplifies calculations tremendously. In this limit, the hopping is scaled
as $T_{nn} = \frac{t^*}{2\sqrt d}$ ($d$ is the space dimension and $t^*$ is the hopping energy unit), the density of states
becomes a Gaussian
\begin{equation}\label{rho0}
\rho_0(\e)= \frac{1}{t^*\sqrt\pi}\exp\lp{-\frac{\e^2}{t^{*2}}}\rp
\end{equation}
and the self-energy
becomes local~\cite{mhart1}
\begin{equation}\label{sis}
\m[\Sigma]_{ij,\sigma}=\m[\Sigma]_{ii,\sigma}\delta_{ij} \ , \quad \m[\Sigma]_{\bk\sigma}=\m[\Sigma]_{ii,\sigma}.
\end{equation}
The spatially uniform electric field
${\bf E}(t)$ is created by the vector potential, aligned along the main diagonal
of a hypercube,
\begin{equation}\label{duzh}
\bA(t) = A(t)(1,1,...) \ .
\end{equation}
The computational scheme is realized as follows. We calculate the Hartree-Fock Green's functions $\m[G]_{0,\bk\sigma}$
using analytic relations. Summation over $\bk$ gives us the local Hartree-Fock Green's functions so that we
can calculate the second-order self-energies from Eqs.~(\ref{sela2}) and (\ref{sis}). Then we solve the Dyson equation in Eq.~(\ref{dmat})
numerically and find the dressed Green's function $\m[G]_{\bk\sigma}$. Equation~(\ref{dmat}) is a linear Volterra
matrix equation of the second kind and allows a very efficient numerical integration~\cite{eck-volterra}. The solution obtained this
way will be the SOPT solution. If we want to obtain SCSOPT solution, we use the SOPT Green's function $\m[G]_{\bk\sigma}$
as a first approximation. Then summing over $\bk$ we obtain a local dressed Green's function $\m[G]_{ii,\sigma}$,
use it to calculate new values for the self-energies in Eqs.~(\ref{sela2}) and (\ref{sis}) and repeat all the previous steps until
the dressed Green's functions converge. Then the lesser Green's function $G^{<}_{\bk\sigma}(t,t')$ is used to find the
charge current, the total energy and the double occupancy according to Eqs.~(\ref{jcur}--\ref{etot}).

This approach can also be generalized to IPT. In that case, one needs to write additional equations for the impurity
model and the self-energy diagram in Fig.~\ref{secd} is applied to the impurity Hamiltonian instead of
the lattice one. We will not discuss the details of the IPT scheme here and instead refer the reader
to articles by Amaricci, {\it et. al.}~\cite{amari} and  Eckstein and Werner~\cite{meck}. Our approach is different
from those only in the absense of imaginary time piece of the contour, which is always valid if one starts from
the noninteracting thermal state.

\section{Numerical results} 
We present the results of SOPT, SCSOPT and IPT calculations of the nonequilibrium current,
double occupancy, total energy and Green's functions as functions of time for a half-filled metal. In equilibrium, perturbative
approaches provide reliable results when $U$ is far from metal-insulator transition, which happens
at~\cite{bulla} $U_c(T=0) \approx 4.1$.
The initial state of
the system is noninteracting ($U=0$) and is in thermal equilibrium at an initial temperature $T_0$. In all
calculations presented in this article, the electric field ${\bf E}$ and Hubbard $U$ are turned on simultaneously
at $t=0$, thus one should remember that temperatures $T_0$ given for each plot characterize only the initial total
energy of the system, since an additional energy of ${\frac{\scriptstyle 1}{\scriptstyle 4}}U$ is instantaneously pumped into the
system due to the sudden change of $U$
at $t=0$ and then subsequent Joule heating can further change the energy when a current flows.
On the plots,
the energy and temperature are measured in units of hopping energy $t^*$, time -- in units of $\hbar/t^*$,
the electric field -- in units of $t^*/(ea)$, where $a$ is the lattice constant, and the current
density -- in units of $et^*/(\hbar a^{d-1})$.

\subsection{Constant electric field}
\begin{figure}[!htb]
\begin{center}
\includegraphics[width=1.\linewidth]{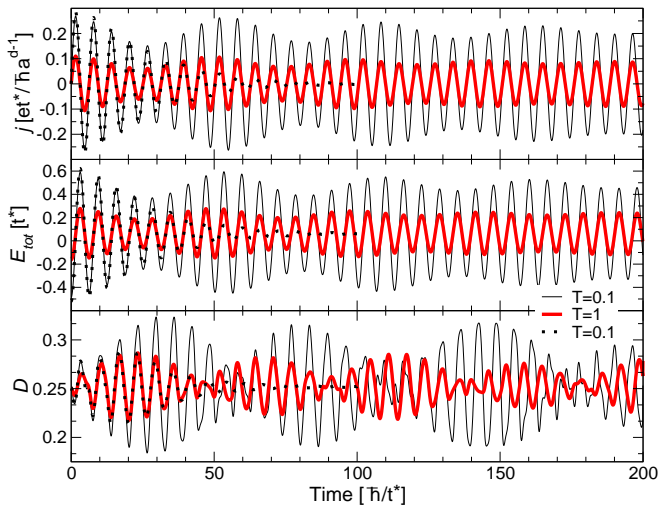}
\end{center}
\caption{(color online) Nonequilibrium (a) current density, (b) total energy and (c) double occupancy
versus time for different values of $T_0$ with $E=1$, $U=0.25$. Solid black lines correspond the SOPT with $T_0=0.1$, solid bold red lines correspond to the SOPT with $T_0=1$, and dotted lines
to the IPT calculation with $T_0=0.1$.}
\label{fig1}
\end{figure}

In Fig.~\ref{fig1}, we plot the current density, total energy and double occupancy for $T_0=0.1$ (black, SOPT
calculation), $T_0=1$ (bold red, SOPT calculation) and $T_0=0.1$ (black dotted, IPT calculation). The system is
a metal ($U=0.25$) placed in a diagonal electric field with an amplitude $A(t)=-E\theta(t)$ and $E=1$.

The SOPT current density (upper panel, solid lines) is a sine function with period $2\pi\hbar/(eaE)$ (or $2\pi/E$ in
the units used for plotting), modulated by a time-dependent amplitude.
The main oscillation period, as well as the physical mechanism of the origin of the oscillating current,
is the same as for the Bloch oscillations in the noninteracting case~\cite{jkf0}.
The modulation produces beats, which become smaller with time,
however do not vanish up to the largest times we have studied, which is $t_{max}=400$; the period of the beats
decreases as the initial temperature increases, which can be seen most easily on the double occupancy panel. Higher initial
temperatures also reduce the amplitude of the current
density oscillations. The IPT curve ($T_0=0.1$, dotted lines) shows a very different behavior.
Up to $t\approx30$ the current coincides with the SOPT result, but then the oscillation amplitude continues
to decline, while the SOPT amplitude shows beats.

The current density plots in Fig.~\ref{fig1} show a dependence of the system behavior
on its initial temperature, despite the fact that additional energy is pumped into the system by the
electric field. The total energy oscillations (middle panel) have a phase shift of $\pi/2$ with respect to the
current density, in agreement with the relationship $dE_{tot}(t)/dt = \bE\cdot{\bf j}$. This
relationship reflects the fact, that the system energy increases, whenever there is a current ${\bf j}$ in the direction
of the electric field $\bE$, and decreases, when the direction of the current is opposite.
The double occupancy (lower panel) is
oscillating close to the noninteracting value of $0.25$. The main oscillation period is the same as for
the current density, but a slow temperature-dependent modulation is now present in both the amplitude and
the phase of the oscillations. Note that for the half-filled model in thermal equilibrium, the repulsive $U$ ({\it i.~e.}~$U>0$)
always leads to a double occupancy tht satisfies $D<0.25$, while values $D>0.25$ are possible only for attractive $U$.
Since we consider only a repulsive interaction here, the transient times where the double occupancy satisfies $D>0.25$ should be viewed as a signature of
nonequilibrium effects.

\begin{figure}[!htb]
\includegraphics[width=1.\linewidth]{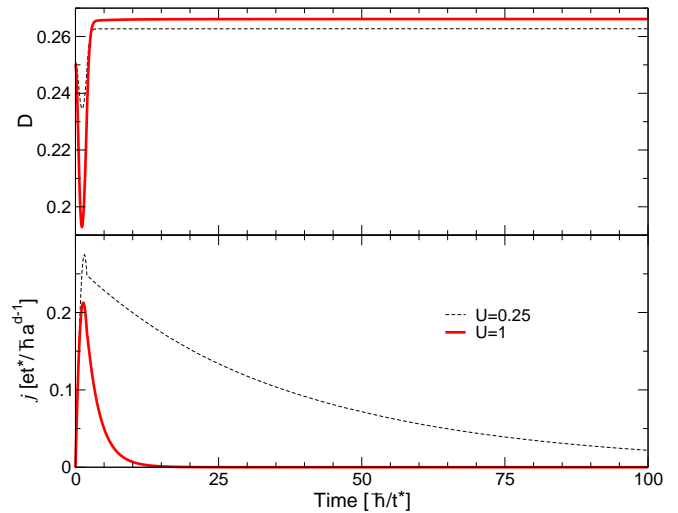}
\caption{(color online) Nonequilibrium (a) double occupancy and (b) current density versus time in the SCSOPT for
different values of $U$ a pulsed field with $E=1$, $T_0=0.1$, $t_f=2$.}
\label{fig5a}
\end{figure}

On the other hand, calculations within the IPT (Fig.~\ref{fig1}, dotted curves) show that in the long time
limit the system always approaches the thermal equilibrium state with $T=\infty$: the current vanishes,
the total energy and the double occupancy approach $E_{tot}=0$ and $D=0.25$ respectively, and
the single particle distribution function (not plotted here) shows that all available states are occupied with the same
probability~\cite{mypriv}. This scenario of heating the system to an infinite temperature is discussed
in Refs.~\onlinecite{prelov} and \onlinecite{meck}.

The SCSOPT calculations (not shown in Fig.~\ref{fig1}) agree well with the IPT results both qualitatively and
quantitatively: current, total energy and the double occupancy coincide within about 5\% up to $t\approx 50$,
but for larger times IPT results show a faster decay. Thus we can conclude that in presence of the electric field, the
SOPT is reliable for small time scales only, unlike the equilibrium case, where the precision of the SOPT
depends solely on the value of the Coulomb interaction $U$.

\subsection{Pulsed electric field}
In this subsection, we examine the case, when the electric field $E$ is acting only within a finite time
interval $t\in[0,t_{f}]$, $E(t)=-E\theta(t)\theta(t_f-t)$. For all the cases studied below, we have
verified that the IPT and the SCSOPT produce similar results, so we choose the latter, since it is less
computationally demanding, therefore we can do calculations using a finer energy grid for higher accuracy.

In Fig.~\ref{fig5a}, we plot the double occupancy and current density for $U=0.25$ (black) and $U=1$ (bold red)
for $E=1$, $T_0=0.1$, and $t_f=2$ calculated within SCSOPT.
Once the electric field is turned off, the double occupancy relaxes to a constant value after a
characteristic time $\approx 1$, which is consistent with an estimate $\tau_{e}\approx\frac{\hbar}{t^*}$
of the time between electron-electron collisions in the half-filled tight binding model. The current
decay can be fit by $j(t)\approx j(t_f)\exp(-U^2(t-t_f)/2)$, thus the characteristic time for current decay turns out to be inversely
proportional to $U^2$ in accordance with Boltzmann equation results~\cite{meck}.

This vanishing of the current after the electric field is turned off is due to the noncommutativity
of the Hamiltonian in Eq.~(\ref{ham}) with the charge current operator in Eq.~(\ref{jop}). Thus the simultaneous presence of
both the lattice and the Coulomb interaction provides another mechanism for breaking quasi-momentum
conservation, in addition to introducing a thermal bath, as suggested in Ref.~\onlinecite{amari}.

It is interesting to note that for small $U$ and short electric field pulses the SOPT produces results which are
almost identical to those of the SCSOPT. Thus the SOPT is accurate for short times, and only fails when the electric field
is present for longer times.

The current and the double occupancy behavior suggest %%give us reasons to believe
that at $t=100$ for both, $U=0.25$ and $U=1$, the system reaches some stationary state. Notice that the
final values of the double occupancy are larger than $0.25$ and increase with the increase of $U$, which is
impossible in thermal equilibrium for repulsive $U$.

\begin{figure}[!htb]
\includegraphics[width=1.\linewidth]{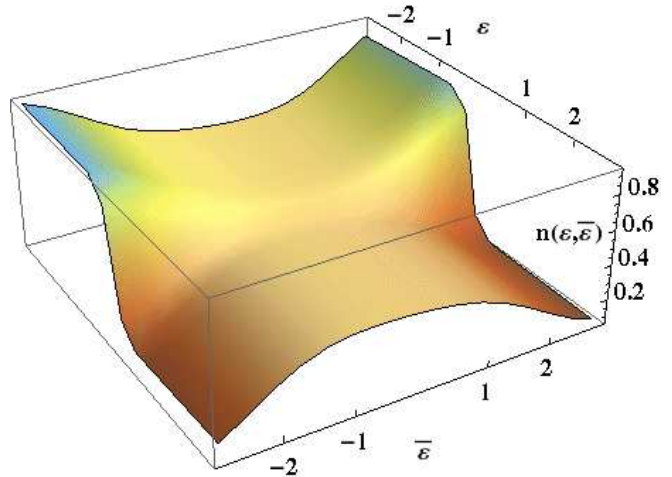}
\caption{(color online) Particle distribution function in the SCSOPT at $t=100$ for a pulsed field with  $E=1$, $U=0.25$, $T_0=0.1$, $t_f=1.625$.}
\label{fig5}
\end{figure}

In order to investigate the origin of this phenomenon, we plot the single-particle distribution function
$n(\e,\bare;t)=-iG^<_{\e,\bare}(t,t)$. As discussed in Ref.~\onlinecite{jkfnonint}, when the ${\bf E}$-field is aligned along the main diagonal of
the hypercube, the $\bk$-dependence of the Green's functions can be reduced to the dependence on two
``energies'' $\e_\bk$ and $\bare_\bk$, compared to the case without electric field, where we have the
dependence of all functions only on $\e_\bk$.
In Fig.~\ref{fig5}, we plot $n(\e,\bare;t=100)$ for $E=1$, $U=0.25$, $T_0=0.1$, and $t_f=1.625$.
If the system has reached a thermal
equilibrium, the resulting distribution would have been independent of $\bare$, while along the $\e$-axis we would
have a thermal distribution, corresponding to some temperature $T$ (and perhaps broadened due to the interactions). Instead, we have a strong $\bare$-dependence
and the $\e$-dependence is far from thermal, especially for $\bare\approx0$, where $n(\e,\bare)$ has a jump
at the Fermi energy $\e=0$, but does not vanish for large $\e$'s. Therefore at large times the system
is stuck in a quasistationary nonequilibrium state. It is exactly this nonequilibrium state, which is responsible for
values of the double occupancies which are incompatible with a thermal state.

\begin{figure}[!htb]
\includegraphics[width=1.\linewidth]{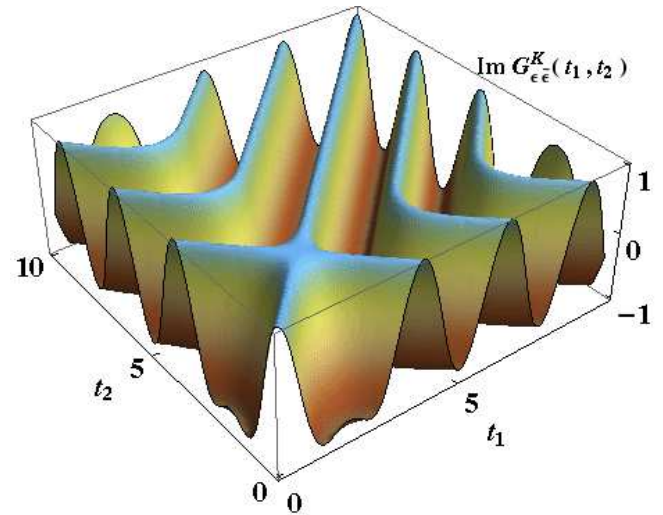}
\caption{(color online) ${\rm Im} G^K_{\e\bare}(t_1,t_2)$ as a function of two times $t_1,t_2$
for the pulsed field with $E=1$, $U=0.25$, $T_0=0.1$, $t_f=3$, and with fixed $\e=-2.625$ and  $\bare=1.125$.}
\label{fig6}
\end{figure}

In Fig.~\ref{fig6}, we plot the imaginary part of the Keldysh Green's function
${\rm Im} G^K_{\e\bare}(t_1,t_2)$ as a function of times $t_1,t_2$ for a pulsed field with $E=1$, $U=0.25$,
$T_0=0.1$, $t_f=3$, and with fixed $\e=-2.625$ and $\bare=1.125$.
We see that whenever $t_1\lessapprox 5$ or $t_2\lessapprox 5$ the Keldysh function
depends on both the average, $t_a=(t_1+t_2)/2$, and the relative, $t_r=t_1-t_2$, times. But when
$t_1\gtrapprox 5$ and $t_2\gtrapprox 5$ this function becomes independent of $t_a$ ({\it i.~e.}~the height of the surface
in Fig.~\ref{fig6} does not change as we move along lines $t_2 = t_1+{\rm const.}$). The same holds for
the real part of the Keldysh Green's function, as well as for the real and imaginary parts of the retarded Green's
function for all values $(\e,\bare)$. Thus we see that the nonequilibrium states of the system, created
by a pulsed $E$-field are so called steady states, {\it  i.~e.}~states independent of the average time.

We can summarize our findings by saying that using an electric field pulse we force a system into a
long-lived nonequilibrium steady state. It was argued by Moeckel and Kehrein~\cite{mokel} that for
small values of the Coulomb interaction, the full thermalization happens on timescales of the order of
$\tau_{therm}\sim \rho_0^{-3}(0)U^{-4}$, so by long-lived we mean that this nonequilibrium steady state does not
thermalize even on this time scale.

\begin{figure}[!htb]
\includegraphics[width=1.\linewidth]{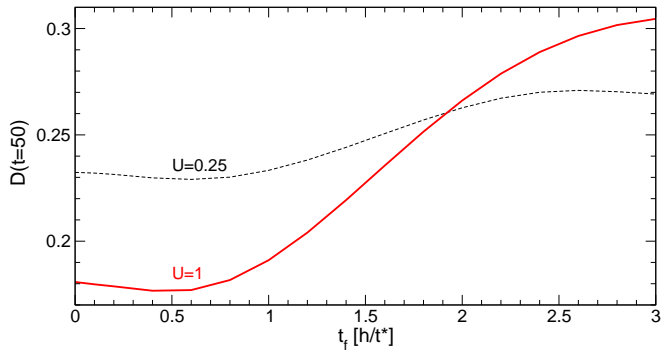}
\caption{(color online) Double occupancy versus the length of the E-pulse $t_f$ in the SCSOPT measured at $t=50$
for $E=1$, $T_0=0.1$ (dashed black) and $T_0=1$ (bold red) and different values of $U$.}
\label{fig7}
\end{figure}

In order to investigate, how the double occupancy of the resulting nonequilibrium steady state depends
on the length of the E-field pulse $t_f$, we plot the double occupancy measured at $t=50\hbar/t^*$ versus $t_f$ in Fig.~\ref{fig7}. 
When $t_f=0$, the double occupancy $D(t=50)$ corresponds to some equilibrium thermal state and is always
less than $0.25$. Depending on the length of the pulse, we can achieve
 $D\approx 0.27$ for $T_0=0.1$ and $U=0.25$. As we can see from Fig.~\ref{fig5a} even
higher double occupancies can be achieved for $U=1$.

\section{Conclusions} 
We have shown how various properties of a nonequilibrium state of the Hubbard
model in a spatially uniform electric field can be calculated perturbatively in the
Coulomb interaction $U$. Such calculations, while being not very computationally intensive,
allow us to access times much longer than other existing methods, so that even a transition of the system to
the steady state can be studied. We have shown that when the interacting system without the external bath
in the metallic state is placed into a DC electric field the Bloch oscillations of the current are
suppressed by the heating of the system to infinite temperature.

We have also shown that a short electric field pulse can create a steady ({\it i.~e.}~average time independent)
nonthermal state, which can exist for times longer than the available theoretical estimates of lifetimes
for nonthermal states.

One might think that the presence of this steady state is just an artifact of the truncation of the perturbation series.  Indeed, at strong couling this can occur if the interaction strength is much larger than the hopping, because relaxation processes require multiparticle effects.  But here we have a weak $U$ and the perturbation theory is a self-consistent one, so it includes many high-order diagrams.  Hence, it is unlikely that these effects arise solely from the truncation of the perturbation theory.
For large $U$, there is evidence that steady states can occur even for continuously driven systems~\cite{fotso}, but the scenario for thermalization is more complex than what is seen here at weak coupling.

\begin{acknowledgments}
We would like to thank V.~Turkowski, O.~Peil, A.~Shvaika, M.~Eckstein,
P.~Arseev and E.~Altman for fruitful discussions. A.J. and A.L. acknowledge support from DFG SFB-925 and CUI-cluster of excellence.
Initial stages of this work was done at Universit\"at Augsburg and many discussions with Prof. D. Vollhardt, as well as support from DFG
Sonderforschungsbereich 484, are acknowledged. J.K.F. acknowledges the National Science Foundation under grant number DMR-1006605 and the
McDevitt bequest at Georgetown University.
\end{acknowledgments}

\end{document}